# A Statistical Analysis of Bowling Performance in Cricket

**Akash Malhotra** [a,1]**, Shailesh Krishna**[b]

[a] *Indian Institute of Technology, Bombay, India*
[b] *Indian Institute of Technology, Bombay, India*

## Abstract

*There is a widespread notion in cricketing world that with increasing pace the performance of a bowler improves. Additionally, many commentators believe lower order batters to be more vulnerable to pace. The present study puts these two ubiquitous notions under test by statistically analysing the differences in performance of bowlers from three subpopulations based on average release velocities. Results from one-way ANOVA reveal faster bowlers to be performing better, in terms of Average and Strike-rate, but no significant differences in the case of Economy rate and CBR. Lower and Middle order batsmen were found to be more vulnerable against faster bowling. However, there was no statistically significant difference in performance of Fast and Fast-Medium bowlers against a top-order batter.*

---
[1] *Corresponding author at:* Indian Institute of Technology, Bombay, Mumbai-400076, India.
*E-mail addresses:* akash_malhotra@iitb.ac.in (A. Malhotra), shailesh.krishna@iitb.ac.in (S. Krishna).

# 1. Introduction

Cricket, as a sport, derives its thrill from the battle between bat and bowl. An efficient bowling unit is every captain's dream which could run through the opponent's batting line-up. Bowling with a cricket ball is a perfect fusion of art and science wherein the bowler, limited by his own biomechanics, exploits the physics governing motion of ball through air as well as deviation off the pitch surface to playing tricks on batter's mental template of the ball's trajectory. This template is an attempt by batter's mind to predict the trajectories of the following deliveries based on release velocity and angle of preceding deliveries. Limited by his reflexes, the batter has to depend more on this mental template as the release velocity of ball increases, leaving the batsman with shortened response time. Consequently, a ubiquitous notion exists in the cricketing world that batsmen find balls released at higher velocities difficult to play. Naturally, this has led cricket enthusiasts to believe that bowlers who release ball at higher velocities tend to outperform those releasing at lower velocities. Lower order batters, generally, do not specialize in batting techniques; based on this assumption many commentators believe lower order batters to be more vulnerable to pace, presenting them as defenceless hunt before faster bowlers.

The present study attempts to justify/debunk these ubiquitous notions existing in cricketing world by statistically analysing the performance of pace bowlers independently in three groups based on range of release velocities. The statistical variation in number of wickets taken by bowlers of different categories are studied for each stratum of batting line-up. The rest of the paper is organised as follows: section 2 explains the aerodynamics behind possible deviations in trajectory of ball arising out of variations in release. Section 3 describes the data and statistical methodology employed in this study. Section 4 discusses the results of statistical analyses and section 5 concludes.

# 2. Release velocity and dynamics of a cricket ball

The physical construction of cricket ball and variations in trajectory arising out of it every time a ball is released from a bowler's hand or bounces off the pitch surface is unique to cricket. Generally, two pieces of leather are stitched together to form the outer covering of a cricket ball. The primary *'seam'* is made up of six rows of stitching bulging out of the equator. This seam is the primary weapon of a bowler using which a bowler makes the ball move laterally in the air, also known as *'swing'*. This is caused by breaking of laminar boundary layer into turbulence at the position of seam on one side of the ball. This creates a pressure differential across the ball generating a side force on the ball as depicted in Fig. 1. Consequently, a ball released with the seam pointing towards *slip* fielders will move away from the batter (*outswing*) and if released with seam angled towards *fine leg* will swing into the batsman (*inswing*). This type of swing bowling is commonly known as conventional swing.

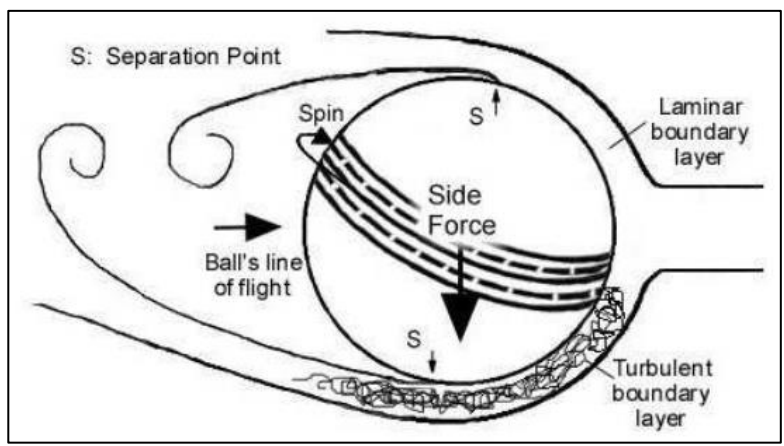

**Figure 1.** Flow over a cricket ball (*Source*: Mehta, 2014)

The maximum sideways force attainable in a range of release velocity is dependent on angle of seam to the initial line of flight (refer Mehta, 2000). The direction of movement (*conventional, reverse* or *contrast*) achieved through swing bowling largely depends on condition of the ball including surface roughness and seam condition (Mehta, 2014). Apart from swing movements, bowlers do exploit *Magnus effect* as well, with majority of pace bowlers imparting backspin to the ball about a near horizontal axis at the time of release producing an upward Magnus force (Fig. 2A). Side-arm bowlers tilts the axis of spin, producing a lateral component of Magnus force which pushes the ball sideways (Fig. 2B).

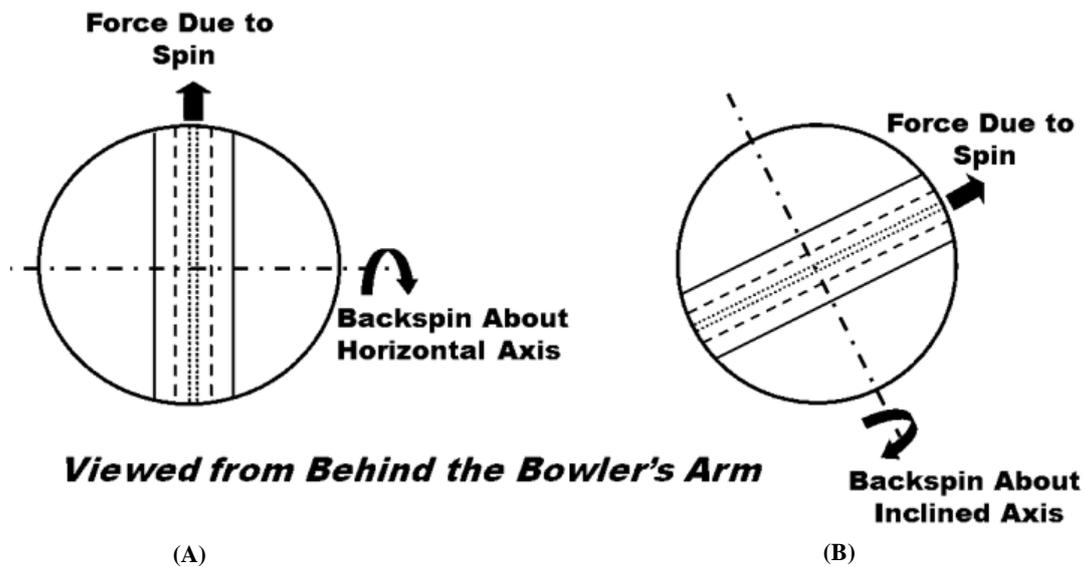

**Figure 2.** Schematic for *'Malinga'* swing (*Source*: Mehta, 2014)

## 3. Statistical Framework

The performance of pace bowlers is studied by categorizing them into three subpopulations based on their average release velocities viz.

i. *Fast* (>142 km/h),
ii. *Fast-Medium* (130-141 km/h), and
iii. *Medium-Fast* (120-129 km/h)

*Spinners*, or particularly bowlers who release ball at velocities below 120 km/h, on an average, are not included in the present study. The indicators used to quantify performance of bowlers in this study include:

Av : Career average of a bowler, computed as ratio of runs conceded per wickets taken
SR : Career strike rate of a bowler, computed as ratio of number of balls bowled per wickets taken
ER : Career Economy rate of a bowler, computed as ratio of runs conceded per *overs*[2] bowled
CBR : Career combined bowling rate (Lemmer, 2002), computed as harmonic mean of Average, Strike rate and Economy rate.

$$CBR = \frac{3}{\frac{1}{Av} + \frac{1}{SR} + \frac{1}{ER}}$$

---

[2] In cricket, over is a set of six balls bowled consecutively by a single bowler from one end of a cricket pitch.

The lower these indicators are in value, the better the bowler is performing.

## 3.1 Data

The bowling record of all the players till December 25$^{th}$, 2016, who played more than 5 international *test* matches, were imported from archives of Howstat (2016). Players who bowl more than 15 overs in a match, on an average, have been considered as full-time bowlers. As the cricket and its playing conditions have changed significantly over time, only bowlers born after 1$^{st}$ January 1955 have been considered for further analysis. The final dataset contains bowling records of 285 bowlers which were categorized into three groups based on their average bowling speed; leaving 62 bowlers in Fast, 168 in Medium-Fast (MF) and 55 in Fast-Medium (FM) category. Apart from the usual cricket statistics, the imported dataset also contains share of wickets taken across the batting order by each bowler, wherein the batting order is divided into three strata viz., Top (1-3), Middle (4-7) and Lower (8-10) order.

## 3.2 Methodology

A one-way ANOVA with post hoc tests are used to determine how the bowler performance differed based on speed categories amongst bowlers. The one-way ANOVA tests whether group means are statistically different from each other. A one-way ANOVA controls the Type I error rate whereas running multiple *t-tests* on all group means in presence of three or more groups causes Type I error to increase unacceptably. The assumption of homogeneity of variances in one-way ANOVA is being tested using Levene's test of equality of variances (Levene, 1960), which determines whether the variances between groups for the dependent variable are equal. The hypotheses for this test could be stated as:

$$H_0 : \sigma^2_{Fast} = \sigma^2_{FM} = \sigma^2_{MF}$$

$H_A$: not all variances are equal in the population

If null hypothesis from Levene's test can't be rejected, then comes one-way ANOVA with hypotheses:

$$H_0 : \mu_{Fast} = \mu_{FM} = \mu_{MF}$$

$H_A$: at least one group population mean is different

Being an omnibus test statistic, the one-way ANOVA, on rejection of null hypothesis, is incapable of telling which specific groups are significantly different from each other. Keppel & Wickens (2004) recommends running a post-hoc test after a statistically significant one-way ANOVA to identify where the group differences specifically lie. Under the assumption of homogeneity of variances between groups, the Turkey post hoc test is useful in in providing the statistical significance level for each pairwise comparison (Westfall et al., 2011). However, the original Turkey test is not suitable for groups with different sample sizes as it was developed for balanced designs. A modified version, called the Turkey-Kramer post hoc test, is being run here to allow for unequal number of samples in the three bowling categories (Hayter, 1984).

If there exists heterogeneity of variances among groups implied by rejection of null hypothesis from Levene's test, the results from standard one-way ANOVA could not be interpreted. Liz et al. (1996) suggests Welch ANOVA as best alternative under violation of variance homogeneity assumption. Under heterogeneity of variances and unequal sample sizes, Ruxton and Beauchamp (2008) recommends using Games Howell post hoc test for pairwise comparisons.

## 4. Results and Discussion

The three groups of independent variable based on bowling speed remains the same for all analyses. With Average (Av) as dependent variable, homogeneity of variances was violated, as assessed by Levene's Test ($p = 0.004$). Accordingly, Welch's F-statistic ($p = 0.000$) reveals bowling averages to be statistically significantly different for the three bowling category groups (see Table 1).

**Table 1** Welch ANOVA for Average ($Av$)

|  | Statistic* | df1 | df2 | Sig. |
|---|---|---|---|---|
| Welch | 11.897 | 2 | 110.700 | 0.000 |
| * Asymptotically F distributed. | | | | |

**Table 2** Multiple Comparisons with Games-Howell for Average ($Av$)

| Games-Howell; Dependent Variable: Average | | | | | | |
|---|---|---|---|---|---|---|
| | | Mean Difference | | | 95% Confidence Interval | |
| (I) Speed | (J) Speed | (I-J) | Std. Error | Sig. | Lower Bound | Upper Bound |
| fast | FM | -4.51402* | 1.45628 | 0.007 | -7.9710 | -1.0571 |
|  | MF | -10.49997* | 2.19167 | 0.000 | -15.7175 | -5.2824 |
| FM | fast | 4.51402* | 1.45628 | 0.007 | 1.0571 | 7.9710 |
|  | MF | -5.98596* | 1.99153 | 0.010 | -10.7467 | -1.2252 |
| MF | fast | 10.49997* | 2.19167 | 0.000 | 5.2824 | 15.7175 |
|  | FM | 5.98596* | 1.99153 | 0.010 | 1.2252 | 10.7467 |
| * The mean difference is significant at the 0.05 level. | | | | | | |

Bowling average increased from the Fast bowling (32.08 ± 9.58) to the Fast Medium (36.59 ± 10.38) to Medium Fast (42.58 ± 13.52) in that order. Games-Howell post hoc analysis (Table 2) revealed that the increase from Fast to FM (4.51, 95% CI (1.06 to 7.97)) was statistically significant ($p = 0.007$), the increase from Fast to MF (10.50, 95% CI (5.28 to 15.72), $p = 0.000$), as well as increase from FM to MF (5.99, 95% CI (1.23 to 10.75), $p = 0.010$).

The homogeneity of variances was violated in case of Strike Rate ($SR$), as assessed by Levene's Test ($p = 0.000$). Subsequently, Welch's F-statistic ($F(2, 113.35) = 24.229$, $p < 0.05$) indicated Strike rate to be statistically significantly different amongst the three bowling category groups (see Table 3). Strike-rate of bowlers increased from the Fast bowling (58.58 ± 13.10) to the Fast Medium(FM) (70.02 ± 17.15) to Medium Fast (81.52 ± 25.44) in that order. Games-Howell post hoc analysis (Table 4) revealed that the increase from Fast to FM (11.44, 95% CI (6.41 to 16.48)) was statistically significant ($p = 0.000$), the increase from Fast to MF (22.94, 95% CI (13.83 to 32.05), $p = 0.000$), as well as increase from FM to MF (11.5, 95% CI (2.7 to 20.30), $p = 0.007$).

**Table 3** Welch ANOVA for Strike-rate ($SR$)

|  | Statistic* | df1 | df2 | Sig. |
|---|---|---|---|---|
| Welch | 24.229 | 2 | 113.346 | 0.000 |
| * Asymptotically F distributed. | | | | |

**Table 4** Multiple Comparisons with Games-Howell for Strike-rate (*SR*)

| Games-Howell; Dependent Variable: Strike Rate | | | | | | |
|---|---|---|---|---|---|---|
| | | Mean Difference | | | 95% Confidence Interval | |
| (I) Type | (J) Type | (I-J) | Std. Error | Sig. | Lower Bound | Upper Bound |
| fast | FM | -11.44072* | 2.12565 | 0.000 | -16.4755 | -6.4060 |
| | MF | -22.94183* | 3.81212 | 0.000 | -32.0487 | -13.8349 |
| FM | fast | 11.44072* | 2.12565 | 0.000 | 6.4060 | 16.4755 |
| | MF | -11.50111* | 3.67609 | 0.007 | -20.3017 | -2.7005 |
| MF | fast | 22.94183* | 3.81212 | 0.000 | 13.8349 | 32.0487 |
| | FM | 11.50111* | 3.67609 | 0.007 | 2.7005 | 20.3017 |
| * The mean difference is significant at the 0.05 level. | | | | | | |

As assessed by Levene's test ($p = 0.207$), there exists homogeneity of variances with Economy rate (*ER*) as dependent variable. The differences in Economy rate between the three bowling groups were not statistically significant ($p = .0150$) as revealed by one-way ANOVA (see Table 5). Similar results are obtained for Combine Bowling Rate (*CBR*), wherein a Levene's test ($p = 0.158$) followed with one-way ANOVA ($F(2, 282) = 0.675$, $p = 0.510$) finds differences in *CBR* amongst the three bowling categories to be statistically insignificant (Table 6).

**Table 5** One-way ANOVA for Economy rate (*ER*)

| | Sum of Squares | df | Mean Square | F | Sig. |
|---|---|---|---|---|---|
| Between Groups | .732 | 2 | .366 | 1.908 | .150 |
| Within Groups | 54.136 | 282 | .192 | | |
| Total | 54.869 | 284 | | | |

**Table 6** One-way ANOVA for Combined Bowling rate (*CBR*)

| | Sum of Squares | df | Mean Square | F | Sig. |
|---|---|---|---|---|---|
| Between Groups | 1.677 | 2 | .838 | .675 | .510 |
| Within Groups | 349.978 | 282 | 1.241 | | |
| Total | 351.655 | 284 | | | |

For every bowler, the total number of wickets taken individually against batters from the three strata are normalized by number of matches played by a bowler. This is done to allow for unbiased comparison between bowler groups within a stratum. With dependent variable as number of top order wickets taken per match by a bowler, there exists homogeneity of variances among bowling groups as assessed by Levene's test ($p = 0.188$). Subsequently, one-way ANOVA ($p = 0.000$) finds statistically significant differences between the three bowling categories (Table 7).

**Table 7** One-way ANOVA for Top order (1-3) wickets per match

| | Sum of Squares | df | Mean Square | F | Sig. |
|---|---|---|---|---|---|
| Between Groups | 2.695 | 2 | 1.348 | 9.444 | .000 |
| Within Groups | 40.242 | 282 | .143 | | |
| Total | 42.937 | 284 | | | |

**Table 8** Multiple Comparisons with Turkey-Kramer for Top order wickets/match

| Turkey-Kramer; Dependent Variable: Top order wickets/match | | | | | | |
|---|---|---|---|---|---|---|
| | | Mean | | | 95% Confidence Interval | |
| (I) Type | (J) Type | Difference (I-J) | Std. Error | Sig. | Lower Bound | Upper Bound |
| fast | FM | .0974522751 | .0561342972 | .194 | -.0348089161 | .2297134665 |
| | MF | .2970956412 | .0699730168 | .000 | .1322282601 | .4619630224 |
| FM | fast | -.0974522751 | .0561342972 | .194 | -.2297134665 | .0348089161 |
| | MF | .1996433661 | .0586855667 | .002 | .0613709842 | .3379157479 |
| MF | fast | -.2970956412* | .0699730168 | .000 | -.4619630224 | -.1322282601 |
| | FM | -.1996433661* | .0586855667 | .002 | -.3379157479 | -.0613709842 |
| * The mean difference is significant at the 0.05 level. | | | | | | |

Top order wickets/match score decreased from the fast (1.14 ± 0.33), to FM (1.05 ± 0.37), to MF (0.85 ± 0.45). Tukey-Kramer post hoc analysis (see Table 8) revealed that the decrease from fast to MF (0.30, 95% CI (0.13 to 0.46)) was statistically significant ($p = 0.000$), as well as the decrease from FM to MF (0.20, 95% CI (0.06 to 0.34), $p = .002$), but Fast and FM group differences were not statistically significant ($p = 0.194$).

**Table 9** One-way ANOVA for Middle order (4-7) wickets per match

| | Sum of Squares | df | Mean Square | F | Sig. |
|---|---|---|---|---|---|
| Between Groups | 5.381 | 2 | 2.690 | 20.240 | .000 |
| Within Groups | 37.485 | 282 | .133 | | |
| Total | 42.866 | 284 | | | |

**Table 10** Multiple Comparisons with Turkey-Kramer for Middle order wickets/match

| Turkey-Kramer; Dependent Variable: Middle order wickets/match | | | | | | |
|---|---|---|---|---|---|---|
| | | Mean | | | 95% Confidence Interval | |
| (I) Type | (J) Type | Difference (I-J) | Std. Error | Sig. | Lower Bound | Upper Bound |
| fast | FM | .2648447519* | .0541775293 | .000 | .1371940119 | .3924954919 |
| | MF | .4163082495* | .0675338494 | .000 | .2571879287 | .5754285703 |
| FM | fast | -.2648447519 | .0541775293 | .000 | -.3924954919 | -.1371940119 |
| | MF | .1514634975 | .0566398650 | .022 | .0180111093 | .2849158858 |
| MF | fast | -.4163082495 | .0675338494 | .000 | -.5754285703 | -.2571879287 |
| | FM | -.1514634975 | .0566398650 | .022 | -.2849158858 | -.0180111093 |
| * The mean difference is significant at the 0.05 level. | | | | | | |

There was homogeneity of variances for Middle order wickets/match, as assessed by Levene's test ($p = 0.270$). One-way ANOVA ($p = 0.000$) revealed statistically significantly differences between the three bowling groups (see Table 9). Middle order wickets/match decreased from the fast (1.31 ± 0.34), to FM (1.05 ± 0.35), to MF (0.89 ± 0.42). Tukey-Kramer post hoc analysis (Table 10) revealed that the decrease from fast to FM (0.26, 95% CI (0.14 to 0.39)) was statistically significant ($p = 0.000$), as well as the decrease from FM to MF (0.15, 95% CI (0.02 to 0.28), $p = 0.022$), and the decrease from fast to MF (0.42, 95% CI (0.26 to 0.58), $p = 0.000$).

**Table 11** One-way ANOVA for Lower order (8-10) wickets per match

|  | Sum of Squares | df | Mean Square | F | Sig. |
|---|---|---|---|---|---|
| Between Groups | 4.013 | 2 | 2.007 | 18.579 | .000 |
| Within Groups | 30.459 | 282 | .108 |  |  |
| Total | 34.472 | 284 |  |  |  |

**Table 12** Multiple Comparisons with Turkey-Kramer for Lower order wickets/match

| Turkey-Kramer; Dependent Variable: Lower order wickets/match | | | | | | |
|---|---|---|---|---|---|---|
| (I) Type | (J) Type | Mean Difference (I-J) | Std. Error | Sig. | 95% Confidence Interval | |
|  |  |  |  |  | Lower Bound | Upper Bound |
| fast | FM | .2008943948 | .0488366468 | .000 | .0858276101 | .3159611796 |
|  | MF | .3686478740 | .0608762856 | .000 | .2252138156 | .5120819325 |
| FM | Fast | -.2008943948 | .0488366468 | .000 | -.3159611796 | -.0858276101 |
|  | MF | .1677534792 | .0510562426 | .003 | .0474569793 | .2880499791 |
| MF | Fast | -.3686478740 | .0608762856 | .000 | -.5120819325 | -.2252138156 |
|  | FM | -.1677534792 | .0510562426 | .003 | -.2880499791 | -.0474569793 |
| * The mean difference is significant at the 0.05 level. | | | | | | |

As assessed by Levene's test ($p = 0.999$), there exists homogeneity of variances in case of Lower order wickets taken per match by a bowler. Furthermore, there exists statistically significant differences (see Table 11) amongst the three group of bowlers as indicated by one-way ANOVA ($p = 0.000$). Lower order wickets/match score decreased from the fast (0.92 ± 0.33), to FM (0.72 ± 0.33), to MF (0.55 ± 0.32) category. Tukey-Kramer post hoc analysis (Table 12) revealed that the decrease from fast to FM (0.20, 95% CI (0.09 to 0.32)) ($p = 0.000$), and the decrease from FM to MF (0.17, 95% CI (0.05 to 0.29), $p = .003$), as well as the decrease from fast to MF (0.37, 95% CI (0.23 to 0.51), $p = 0.000$) were all statistically significant.

## 5. Conclusions

According to the results from versions of ANOVA on various performance parameters, in terms of average and strike rate, Fast bowlers are, indeed, superior than Fast-Medium and Medium-Fast bowlers, and also, Fast-Medium bowlers outperform Medium-Fast bowlers based on these parameters. An important observation which arises out of analyses made in this study is that even a Fast bowler takes less balls and runs to take a wicket, he still concedes almost the same number of runs in an over as a Fast-Medium or Medium-Fast bowler. A probable reason for this could be that fast bowlers tend to bowl more in the beginning of innings when the red ball is new and has more uncertainty associated with its trajectory; consequently, the batter *edges* the new ball quiet often which sometimes goes in the hands of *wicket-keeper* or *slip* fielders adding to wicket tally and, if not, leaks runs while passing above and between *slip* and *gully* fielders and resulting in *boundaries*. This could also be due to the fact that because of higher release velocities, *mishits* against a Fast bowler tend to go long way, if not *caught*. In pursuit of psychological avoidance to bat against fast bowlers, batsmen many times try to counter-attack and score quickly against Fast bowlers so that the captain removes that particular bowler from bowling attack.

However, bowlers from all the three categories have no statistically significant differences in their performance when measured in terms of CBR, a measure which is by far the most complete and a holistic measure of a bowler's performance. With time, the game of test cricket has registered faster innings run rate and debuts of batsmen with higher strike rates; the last 25 years have witnessed faster *outfields* and significant advancements in protective gear as well as bats used by batters, while the bowling has remained more or less same. Therefore, the statistical comparison of bowler performance across times without adjusting for these changes is not wholly perfect. This is an important note for future analysis of cricketing records. The unreliability of data obtained from internet further limits the accuracy of statistical results presented in this study.

Considering the second notion regarding effectiveness of bowling speeds against batsmen from different positions in batting line-up, there is no statistically significant difference in wickets/match taken by Fast and Fast-Medium bowlers against a top-order batsman; however Medium-fast bowlers lag behind in terms of Top-order wickets/match. Top order (1-3) batsmen tend to be most skilled in the entire batting line-up and, hence can adjust their mental templates to high release velocities of Fast bowlers and might even find slower speeds of Medium-fast bowlers relatively easy to play. Against middle (4-7) and lower (8-10) order batsmen, the succession of wickets taken per match remains as expected: Fast > FM > MF; reinforcing the notion that these batters are, indeed, vulnerable to pace. Interestingly, the analysis revealed that mean of wickets/match for every bowling category has higher values against top order as compared to lower order batsmen even though there are equal number of batters in both the strata[3]. This is probably due to the fact that lower order batsmen do not get a chance to bat at numerous occasions - an innings gets declared, target gets chased with loss of few wickets or a match gets drawn.

## Supplementary Material

The dataset for replication of results could be accessed at http://dx.doi.org/10.7910/DVN/ZJCQP7.

---

[3] Middle order (4-7) comprises of four batters as opposed to top (1-3) and lower (8-10) order which comprises of three batters each.